\documentstyle[12pt,aaspp]{article}
\begin{document}
\title{Rees-Sciama Effect in a CDM Universe}
\author{Uro\v s Seljak\altaffilmark 1}
\affil{Department of Physics, MIT, Cambridge, MA 02139 USA }
\begin{quote}
\altaffilmark{}
\altaffiltext{1}{Also Department of Physics, University of Ljubljana,
Jadranska 19, 61000 Ljubljana, Slovenia}
\end{quote}
\begin{abstract}
The Rees-Sciama (RS) effect produces fluctuations in the cosmic
microwave background (CMB) through the time-dependent gravitational
potential in the nonlinear stages of evolution.
I investigate the RS effect on the CMB angular power spectrum $C_l$
for several CDM models by combining the results of N-body simulations
with second order perturbation theory.
The amplitude of the RS fluctuations peaks at $l \sim 100-300$, where
it gives $\Delta T /T \sim 10^{-7}-10^{-6}$ for a wide range of models.
This is at least an order of magnitude below
the COBE normalized primary contribution. RS fluctuations could
be a dominant source of anisotropies only on subarcminute scales
($l \approx 5000$) and are below the present day observational
sensitivities on all angular scales.

\end{abstract}
\section{Introduction}
It is widely believed that cosmic microwave background anisotopies
observed by COBE (Smoot et al. 1992) and smaller scale experiments are
caused by small inhomogeneities in the matter distribution
during the recombination
epoch (redshift $z \approx 1100$).
At that time the fluctuations were linear and in
principle calculable with arbitrary precision, which would allow one to
determine several cosmological parameters with a very high precision
(e.g. Bond 1995; Hu \& Sugiyama 1995; Seljak 1994).
This picture could
be complicated by the presence of nonlinear contributions to the
anisotropies, arising from the late stages of evolution.
These are determined by different physical
processes and are,
because of the uncertainties in the nonlinear evolution,
difficult
to calculate even in well specified models. The most
important contributions are the so-called Vishniac, Sunyaev-Zeldovich and
Rees-Sciama effects.
The first two effects are caused by Thomson scattering of photons off
the free electrons moving in a bulk and random motion, respectively.
They require an ionized medium and
strongly depend on the ionization history of inter and intra-cluster medium,
determined by the complicated physics of collisional gas (see e.g. Persi et
al. 1995; Bond 1995 and references therein).
The third effect, which is explored in this paper, is
caused by a time dependent gravitational potential during the nonlinear
stages of evolution. It does not depend on the
ionization history of the universe, but only on the evolution of gravitational
potential and is thus less model dependent.

The imprint of nonlinear clustering on the CMB
was first pointed out by Rees \& Sciama in 1968
and has
subsequently been analyzed by several authors.
Most of this previous work gave only partial
answers, studying for example isolated structures, such as clusters,
superclusters and voids (Rees \& Sciama 1968;
Kaiser 1982; Nottale 1984; Thompson \& Vishniac 1987; Panek 1992;
Mart\' inez-Gonz\' alez, Sanz \& Silk 1990; Chodorowski 1992, 1994; Arnau,
Fullana \& S\' aez 1994),
quasi-linear (Mart\' inez-Gonz\' alez, Sanz \& Silk 1992)
or strongly nonlinear regimes (Mart\' inez-Gonz\' alez, Sanz \& Silk
1994).
Recently, Tuluie \& Laguna (1995) presented a detailed N-body analysis of
a standard CDM model using ray-tracing of photons. This approach has the
advantage of
producing real maps of $\Delta T/T$, thereby allowing one to identify
the
non-gaussian features that contribute to the Rees-Sciama effect.
Unfortunately such approach is also computationally expensive and the results
have a rather small dynamic range, in the case of Tuluie \& Laguna (1995) being
limited by the number of traced photons and by the
resolution of their $64^3$ PM simulation. For this reason
these authors only present results on degree angular scales for one
particular model.

The approach presented in this paper similarly uses output of N-body
simulation to calculate the angular power spectrum of Rees-Sciama
effect.
The
effect is calculated from the power spectrum using
positions and velocities of particles in the simulation. This avoids the
need to trace the photons through
a dedicated N-body simulation and so it is not limited by the finite number of
photons. Moreover,
one can use already existing N-body simulations with large spatial
resolution, allowing to calculate the effect with a much larger dynamic range.
The range can be further extended to the
scales larger than the size of simulation by matching the N-body
results with the second order perturbation theory calculation, all
of which allows one to calculate the effect with a high accuracy
over most of the angular range of
observational interest today.
Another advantage of the approach used here is that several
different CDM models can be analyzed
with the same N-body simulation by rescaling
its time and length, which allows one to asses the
sensitivity of the effect to the change in the shape and amplitude
of the power spectrum. In section 2 I present all the necessary formalism to
calculate the RS effect. In section 3 the formalism is applied to several
CDM models, which is followed by discussion of the results and conclusion.

\section{Power Spectrum of Potential Time Derivative}

One can approximate the CMB temperature anisotropy $\Delta T/T
(\vec{n}) \equiv \Delta(\vec{n})$
in the direction $\vec n$
as a contribution from the last scattering surface at recombination
and a line-of-sight integral over the
conformal time $\tau$,
(Sachs \& Wolfe 1966; Mart\' inez-Gonz\' alez et al. 1990),
\begin{equation}
\Delta(\vec{n})=\Delta(\vec{n})_{rec}+\int_{\tau_{rec}}^{\tau_0}
2\dot{\phi}d\tau,
\end{equation}
where $\tau_{rec}$ is the recombination time, $\tau_0$ the present time and
first term
$\Delta(\vec{n})_{rec}$ is the primary contribution to the CMB anisotropy
created at the recombination.
The second term on the right hand side is the
integrated Sachs-Wolfe contribution and depends on the time derivative (with
respect to the conformal time $\tau$) of the
gravitational potential $\phi$ along the line-of-sight.
In the expression above the effect of
Thomson scattering was neglected, which is valid if
the dominant contributions to the integrated Sachs-Wolfe term
come from low redshifts where the universe
is optically thin independent of its ionization state. The
integrated Sachs-Wolfe term and in particular its nonlinear contribution is
usually associated with the RS
effect.
Note that the effect is
frequency independent, because it is caused by the gravitational shifting
of photons. This means that it cannot be separated from
the primary
contribution using a multi-frequency spectral information and the only way to
identify it is to specify its spatial distribution.

The magnitude of the RS effect as a function of angle is studied in terms of
the angular power spectrum $C_l$,
which is defined as the $l$-th Legendre
expansion coefficient of the correlation function,
\begin{equation}
C(\theta)=\langle \Delta(\vec n_1) \Delta(\vec n_2) \rangle_{\vec n_1 \cdot
\vec n_2 =\cos \theta}= \sum_l(2l+1)P_l(\cos \theta)C_l,
\end{equation}
where $P_l$ is the $l$-th Legendre polynomial.
Although for nonlinear processes studied here the
power spectrum is not
a sufficient statistic, it nevertheless provides a useful tool to compare
contributions between various processes
on the same angular scale,
provided that they are statistically uncorrelated. Here I
compare the anisotropies arising
from the Rees-Sciama effect with the primary anisotropies arising at the
recombination. The two contributions are spatially well separated
and can be treated as uncorrelated.
In addition to this there might be other secondary
contributions to CMB anisotropy
$\Delta$, such as the Sunyaev-Zeldovich or Vishniac
effect mentioned above,
both of which are also caused by the clustering of large-scale structure
and are thus not necessary uncorrelated
with RS effect. I will not discuss this general case here, since the main
goal is
to answer the question: can the RS effect dominate over
the primary contribution on a given angular scale? If this indeed turns out
to be the case, then a more detailed study of the RS effect would be needed,
including
the analysis of its higher order moments (e.g. Munshi, Souradeep \& Starobinski
1994;
Mollerach et al. 1995), producing real
sky maps (e.g. Tuluie \& Laguna 1995)
and cross-correlating the RS effect with the other secondary sources that are
important.

Expanding the potential time derivative
$\dot{\phi}$ in the spherical basis and using
orthonormality of spherical harmonics
one obtains the following expression for the multipole moments of
the RS effect in a flat universe (Seljak 1994),
\begin{equation}
C_l=(4\pi)^2\int k^2P_\phi(k)dk
\left[\int_0^{\tau_0}
2\dot{F}(\tau)
j_l(kr)\right]^2
d\tau.
\label{cl}
\end{equation}
$P_\phi(k)$ is the power spectrum of potential,
$j_l(x)$ is the spherical Bessel function, $F(k,\tau)$ is the
growth rate of potential and $r$ is the
comoving distance between the photon
at a conformal time $\tau$
and the observer, $r=\tau_0-\tau$. I assumed the distance to
the last-scattering surface is given by $r \approx \tau_0$.
Equation \ref{cl} is only
valid in the linear regime, because of the assumption that a given mode
is only changing in amplitude and not in phase, which allows
a simple description of time dependence in terms of the growth factor
$F(\tau)$,
which is independent of the wavenumber.
It also assumes that the relative separation between two
photons is not affected by
the gravitational lensing,
a valid assumption on the scales of interest here (e.g. Seljak 1995).

The solution in equation \ref{cl} simplifies considerably if
one is considering small angular scales and if the fluctuations at
widely separated points can be considered statistically independent
(a "fair sample" criterion). The latter condition is satisfied
if the window function is broad compared to the largest correlation
length, as it is the case
in the late epoch of time
dependent gravitational potential.
Moreover, in this regime the
correlations at a distance $k^{-1}$ are
slowly changing on a time scale $(ck)^{-1}$,
because weak field
gravity can only produce nonrelativistic motions.
The radial integral in equation \ref{cl} is thus a product of a spherical
Bessel function $j_l(kr)$ and a slowly changing function of time. This
integral may be approximated by removing the slowly
changing part and using the large $l$ approximation,
\begin{equation}
\int_0^x j_l(x')dx'= \left\{ \begin{array}{ll}0,\ x<l\\
\sqrt{ {\pi \over 2l}},\ x \ge l
\end{array}
\right.
\label{jl}
\end{equation}
Combining equations \ref{cl} and \ref{jl} one obtains
\begin{equation}
C_l^{(RS)}=4(4 \pi)^2\int_0^{\infty}P_{\dot{\phi}}[k,\tau=\tau_0-k/l)]{\pi
\over 2l}
S(k\tau_0-l)dk=
32\pi^3\int_0^{\tau_0}{P_{\dot{\phi}}(l/r,\tau)d\tau \over r^2},
\label{phidot}
\end{equation}
where $S(x)$ is a step function being 0 below $x$ and 1 above it.
I introduced
\begin{equation}
P_{\dot{\phi}}(k,\tau)=\dot{F}^2(k,\tau)P_\phi(k),
\end{equation}
the power spectrum of $\dot{\phi}$.
Equation \ref{phidot} is also valid in an open universe on
small angular scales, provided that $r$ is interpreted as the
angular distance, $r=R\sinh[(\tau_0-\tau)/R]$,
where the curvature $R$ can be expressed as $R=(1-\Omega_{0})^{-1/2}H_0^{-1}$,
where $H_0$ is the Hubble constant today.
The assumption that the power
spectrum of $\dot{\phi}$ is not
changing over a timescale
$(ck)^{-1}$ guarantees the validity of
equation \ref{phidot} both in the
linear and in the
nonlinear regime (where both the amplitude and the phase
of a mode are changing with time and the growth factor $F$ depends
on the wavenumber $k$).
The slow time dependence in $P_{\dot{\phi}}$
may be restored when performing
the radial integral in equation \ref{phidot}.

The power spectrum of $\dot{\phi}$
needs to be specified as a function of time and
scale to compute the RS effect.
The potential is related to the density
through the
Poisson equation, which in Fourier space is given by
\begin{equation}
-k^2\phi={3 \over 2}\Omega_{m0}H_0^2a^{-1}\delta.
\label{poisson}
\end{equation}
Here $\Omega_{m0}$ is the matter density in units of critical density
today,
$\delta$ is the
matter density perturbation and $a$ is the expansion factor normalized
to unity today.
In the linear regime $\dot{F}$ is readily evaluated using
the well known solution for the growing mode of density perturbations
$D_+(\tau)$, $F(\tau) \propto D_+(\tau)/a(\tau)$ independent of $k$.
For the zero curvature
model with
a cosmological constant ($\Omega_{m0}+\Omega_{v0}=1$) the growth factor
is given by
(Heath 1977)
\begin{equation}
D_+(a)={\sqrt{\Omega_{m0}+\Omega_{v0}a^3} \over a^{3/2} }
{\int_0^{a}X^{3/2}da \over \int_0^1 X^{3/2}da },
\label{dvacuum}
\end{equation}
where $X=a/(\Omega_{m0}+\Omega_{v0}a^3)$ and $H_0\tau=\int_0^a da/
(\Omega_{m0}a+\Omega_{v0}a^4)^{1/2}$. In an open universe with no
cosmological constant ($\Omega_{m0}<1$, $\Omega_{v0}=0$)
the growth factor is (Heath 1977)
\begin{equation}
D_+(\tau)=-{3 \sinh(\tau/R)[\sinh(\tau/R)-\tau/R]
\over [\cosh(\tau/R)-1]^2}-2, \ \
a={\Omega_{m0} \over 1-\Omega_{m0}}{\cosh(\tau/R)-1 \over 2},
\label{dopen}
\end{equation}
For closed universe ($\Omega_{m0}>1$)
the solution is obtained by analytic continuation and
for flat case by Taylor expansion of equation \ref{dopen}
in the limit $R \rightarrow \infty$.

In a flat $\Omega_{m0}=1$ universe $D_+(\tau)
\propto
a(\tau)$ and $P_{\dot{\phi}}(k)$ vanishes in the linear regime.
In this case the lowest order contribution arises from the second order
perturbation theory, where the density is expanded into
$\delta=a\delta_1+a^2\delta_2$, which gives
\begin{equation}
\dot{\phi}=-{3 \over 2}{H_0^2\over k^2}\dot{a}\delta_2.
\label{phidot2}
\end{equation}
The power spectrum of $\delta_2$ is given by several authors
(Peebles 1980; Mart\' inez-Gonz\' alez, Sanz \&
Silk 1992; notation of the paper by Jain \& Bertschinger 1994
is used below),
\begin{eqnarray}
%P_{\dot{\phi}}=\left( {3H_0\over 2}\right)^2\left({ H_0 \over k}
%\right)^4aP_{22}(k) \\
P_{22}(k)=\int
d^3qP_\delta(q)P_\delta(|\vec{k}-\vec{q}|)F_2^2(\vec{q},\vec{k}-\vec{q}),\\
F_2(\vec{k_1},\vec{k_2})= {5 \over 7} +
{2 \over 7} \left({\vec{k_1} \cdot \vec{k_2} \over k_1^2 k_2^2 }\right)
+ {\vec{k_1} \cdot \vec{k_2} \over 2} \left( {1 \over k_1^2} + {1 \over k_2^2}
\right),
\end{eqnarray}
where $P_\delta(k)$ is the linear density power spectrum. The power spectrum of
the potential time derivative is then
given by $P_{\dot{\phi}}=9/4(H_0/k)^4\dot{a}^2P_{22}$.

In the fully nonlinear
regime even the second-order perturbation theory breaks down and
the behavior of the power spectrum $P_{\dot{\phi}}(k,\tau)$ as
a function of time becomes more complicated.
It can only be calculated using numerical N-body simulations.
An output from an N-body simulation consists of the
positions and velocities of the particles in the simulation box.
{}From this one can calculate the
(over)density field $\delta(\vec{r})=\rho/\bar{\rho}-1$
and momentum density field $\vec{p}(\vec{r})=(1+\delta)\vec{v}$
on a fixed grid in the box by counting the
number of particles and their velocities near each grid point.
Fourier transformation of these quantities gives
$\delta(\vec{k})$ and $\vec{p}(\vec k)$; for simplicity I will drop their
explicit $k$-dependence in the following.
Taking the time derivative of the
Poisson equation \ref{poisson}
and using the continuity equation
\begin{equation}
\dot{\delta}+i\vec{k}\cdot \vec{p}=0,
\end{equation}
one obtains the following expression,
\begin{equation}
\dot{\phi}={3 \over 2}\left( {H_0 \over k} \right)^2\Omega_{m0}a^{-1}(
\eta\delta+i\vec{k} \cdot \vec{p}),
\label{phidotnl}
\end{equation}
where I introduced $\eta \equiv \dot{a}/a$.
This relation connects the potential time derivative to the
density and momentum density.
By averaging over
all different modes with the same amplitude $k$ one obtains the
power spectrum $P_{\dot{\phi}}$.

An example of various power spectra computed from a high-resolution
simulation of a standard CDM (Gelb \& Bertschinger 1994)
is shown in figure \ref{fig3_1}.
The spectra have been calculated at high enough redshift ($z=4$) to be
still in the linear regime for the long wavelengths and are
multiplied by $4\pi k^3$ to obtain a dimensionless quantity. Dotted line
and dashed-dotted
line show the power spectrum of $\delta$ and $i\eta^{-1}\vec k \cdot
\vec p$,
respectively.
The two spectra agree on large scales,
where the linear theory is a good approximation and gives $\delta \approx
-i\eta^{-1}\vec k \cdot
\vec p$. On smaller scales they
start to deviate from one another with the divergence of momentum density
having more power than the density on the same scale. The time derivative
of potential
$\dot{\phi}$ is proportional to the sum of the two
quantities (equation \ref{phidot2}) and is given by the solid line.
It starts much lower than the
density power spectrum, but eventually rises above it and becomes dominated
by the divergence of momentum density. This shows that it is the motion of the
matter that makes a dominant contribution to $\dot{\phi}$ in the nonlinear
regime.
The dashed line shows the corresponding spectrum from
the second-order perturbation theory
calculation.
On large scales the two spectra agree
well, except at the longest wavelength bin, where the disagreement is
caused both by insuficient sampling of the largest mode
and possibly by the absence of long-wavelength coupling in the N-body
simulation.
%Because there is no power present on scales larger than the size of the
%%simulation
%box,
%N-body simulations neglect the nonlinear coupling of these scales to smaller
%%scales.
%This is particularly important for scales just below the box size,
%where this coupling is most important.
On smaller scales the N-body simulation power spectrum
rises above the corresponding second-order perturbation case and leads to
an increase in the RS effect compared to the second order calculation.

The agreement between the results of the N-body simulation and
second-order perturbation theory as a function of expansion factor $a$
is studied in figure \ref{fig3_2}. The second-order power spectrum grows as
$a^4$ and at late times it eventually rises above the N-body spectrum on
large scales.
On smaller scales the N-body spectrum dominates over
the second-order power spectrum.
For $k<1h$ Mpc$^{-1}$ there is a qualitative agreement between the two
predictions, which gives confidence that one may use the results from
the second order perturbation calculation on scales larger than the
size of simulation box. The discrepancy present at late times even
at the longest wavelengths in the simulation could be caused either by
the nonlinear effects beyond the second order or by the absence of
long-wavelenth
coupling in the simulation. It
leads to some uncertainty in the final results, which are discussed in the
following section.

Another approach used in the literature is to approximate the
evolution of $\dot{\phi}$ using only the
evolution of density (or potential) power spectrum (Mart\' inez-Gonz\' alez et
al. 1994).
For example, one could use
semi-analytic approximations by Hamilton et al.
(1991), which model the evolution of potential power spectrum
and try to deduce the power spectrum of $\dot{\phi}$ using
$P_{\dot{\phi}}=(d(P_\phi)^{1/2}/d\tau)^2$.
This approximation assumes that for a
given mode only its amplitude is changing with time, while its phase remains
constant and is equivalent to the approximation used by Mart\' inez-Gonz\' alez
et al. (1994).
It gives valid results in the linear regime, but
breaks down in the nonlinear regime. This is explicitly shown with the
dashed-dotted curve denoted with MSS in figure
\ref{fig3_1}, where
one can see that it gives a poor agreement with the full treatment both in
the perturbative and in the strongly nonlinear regime.
In the second order perturbation
theory the density power spectrum
receives contributions both from $\langle \delta_2 \delta_2 \rangle$
and from $\langle \delta_1 \delta_3 \rangle$. The two contributions are of
the same magnitude and partially cancel each other (Jain \& Bertschinger
1994), leading to a severe underestimation of $P_{\dot{\phi}}$.
In the strongly nonlinear regime the power spectrum of $\dot{\phi}$
is dominated by the momentum density,
which is determined by the momentum part of the
single particle phase space. Its evolution is faster than the
evolution of the density power spectrum, which is determined solely
by the positions of particles, again leading to an underestimate of
$P_{\dot{\phi}}$. Therefore, even with a correct time evolution of
density power spectrum (which was not used by Mart\' inez-Gonz\' alez
et al. 1994) one cannot obtain a reliable estimate of $P_{\dot{\phi}}$.
For its proper description
one needs to specify the full particle phase space information, given by
both the density and the momentum density fields.

There is another Rees-Sciama
contribution to the CMB anisotropies associated with the
creation of vector metric perturbations.
The effect on CMB is described by the vector component of
the integrated Sachs-Wolfe term,
$
\Delta(\vec{n})=\int_{\tau_{rec}}^{\tau_0}
\vec n \cdot \dot{\vec w}d\tau
$ (Sachs \& Wolfe 1967),
where $\vec w$ is the vector metric perturbation and is
created by the transverse momentum density $\rho \vec v_\perp$
(e.g. Bertschinger 1995),
\begin{equation}
-k^2\vec w=
      16\pi Ga^2\rho\vec v_\perp.
\label{vec}
\end{equation}
In the nonlinear regime vector perturbation is suppressed by
$v/c$ compared to the scalar perturbation,
which has the density $\rho$ as a source
(equation \ref{poisson}).
In the perturbative regime a more careful comparison is needed, because
both contributions vanish in the lowest order.
An estimate of the vector amplitude can be obtained
by taking the time derivative of equation \ref{vec}
and using the Euler's equation for $\rho \vec v$. This gives
$\dot{w} \propto k \phi^2$, which has to be
compared to $\dot{\phi} \propto k^2H_0\phi^2$
from equation \ref{phidot2} (with $\delta_2 \propto \delta_1^2$)
for the scalar contribution.
The vector contribution is thus suppressed by $(kH_0)^{-1} \ll 1$
relative to the scalar contribution
and may safely be neglected as a source of CMB anisotropy
both in the perturbative and in the strongly nonlinear regime.

\section{Angular Power Spectrum of the Rees-Sciama Effect}

The N-body results were obtained from a particle-particle/particle-mesh
simulation of a standard CDM model with $\Omega_{m0}=1$ and
$H_0=50$ km/s/Mpc (Gelb \& Bertschinger 1994). This is
a $(50h^{-1}$Mpc$)^3$ simulation with
$144^3$ particles and a resolution of 32$h^{-1}$kpc, normalized
to linear $\sigma_8=1$ today (i.e. the linear mass overdensity averaged over
spheres
of radius $8h^{-1}$Mpc is unity today). The power spectrum of $\dot{\phi}$ was
calculated on a $384^3$ grid.
No shot-noise subtraction was applied to the results and for this reason only
the
lower half of k-modes were used in the actual analysis.
%At early times ($z>4$)
%even this was not sufficient and another N-body simulation with 8 times more
%particles was used to reduce the shot noise.
The largest mode in the simulation was excluded because of insufficient
sampling and/or large-scale cutoff problems.
This resulted in the
dynamic range of N-body simulation between 0.27 and 25$h$Mpc$^{-1}$ in $k$.
For $k<0.27h$Mpc$^{-1}$ and $z>9$ the second order perturbation theory
calculation was used.

Although the N-body simulation was performed for the standard CDM model,
one can change the
parameters of the model without having to use a different simulation or even
to recalculate the power spectrum
of $\dot{\phi}$.
For example, a change in the normalization amplitude $\sigma_8$ corresponds
to a change in the expansion factor and N-body results at expansion factor $a$
can be used
as N-body results today for a different CDM model with $\sigma_8=a$. Similarly
we
can also rescale the length, which corresponds to a
change in the shape of the CDM power spectrum.
To create a CDM model with $\Omega_{m0} h=0.25$, which is the model that agrees
best with recent large-scale structure surveys
(e.g. Peacock \& Dodds 1994; da Costa et al. 1994), one needs
to rescale the distance
by a factor of 2 and instead of 50$h^{-1}$ Mpc box the size of simulation
becomes 100$h^{-1}$ Mpc.
At the same time the normalization also changes, because $8h^{-1}$Mpc scale
corresponds to a twice smaller scale in the box, which has more
power than the original
$8h^{-1}$Mpc
scale.
In such a model the
output at $a=0.61$ corresponds to today if $\sigma_8=1$. If one adopts
$\sigma_8=0.6$ as suggested by cluster abundances (White, Efstathiou \& Frenk
1993),
then today corresponds to $a=0.36$ in the original simulation.
%and there are only few time outputs available before the N-body power
%spectra are completely swamped by the shot noise.
%One can however still
%make a reliable prediction in the regime where second order results are
%valid, which, as shown in figure \ref{fig??},
%is not very far off the N-body results in
%the interesting regime for $l<???$.

Figure \ref{fig3_4} shows the $C_l$'s for various CDM models discussed above.
In all cases the prediction for the temperature anisotropy $\Delta$
is between $10^{-7}$ and $10^{-6}$ over a large
range of $l$, which is at least an order of magnitude
below the predictions from the
primary anisotropies in standard recombination CMB models.
The RS anisotropies are
sensitive to the normalization and shape
of the power spectrum. To the extend that
the second order theory is valid the amplitude scaling is given
by $\sigma_8^{4}$ and so a change of $\sigma_8$ by a
factor of 2 leads to
more than an order of magnitude effect
in $C_l$'s \footnote[1]{When comparing the RS effect
to the primary anisotropies it
is customary to
normalize the primary contribution to COBE,
which does not change with $\sigma_8$.
Normalizing
it to $\sigma_8$ the ratio between the two spectra scales as
$\sigma_8^2$.}.
For a given $\sigma_8$ a decrease in $\Omega_{m0} h$ gives more
power to the large scales and the Rees-Sciama effect increases on large
angular scales. This is
particularly significant for low values of
$l$ where changing $\Omega_{m0} h$ by a factor
of two leads to almost an order of magnitude effect in $C_l$, but is less
important for $l \approx 1000$, which is dominated by the scales that
contribute to the $\sigma_8$ normalization and
so primarily depends on the amplitude of the power spectrum.
For the model that best fits the present day clustering properties
($\Omega_{m0} h=0.25$ and $\sigma_8=0.6$)
the RS effect peaks at $l \approx 100$, where its power is three orders of
magnitude below the primary signal. Even for the most extreme model
studied here ($\sigma_8=1$ and $\Omega_{m0}h=0.25$)
the RS effect is two orders of magnitude weaker than the
primary signal on degree scales and never exceeds $\Delta T/T \sim 10^{-6}$.
Most models become
significant in comparison
with the primary anisotropies only around $l \approx 5000$ where
$\Delta T/T \sim 10^{-7}$,
well below the present observational sensitivity.
For the standard CDM
model the power spectrum obtained using only the second order calculation is
also plotted on figure \ref{fig3_4}.
It agrees with the full calculation on large scales,
overestimates slightly
the $C_l$'s at intermediate scales ($300<l<3000$) and
underestimates them at high $l$, where strongly nonlinear effects become
dominant. In the regime where the primary anisotropies are important
($l<1000$),
second order calculation gives reliable results and may even overestimate
the anisotropies, contrary to the expectation that it
significantly underestimates them (Mart\' inez-Gonz\' alez
et al. 1994).

Lack of N-body data on large and small scales leads to some uncertainty
in the angular power spectrum. One can see from
figure \ref{fig3_2} that at late times
there is some discrepancy between the second order calculation
and N-body results even at long wavelengths and the extrapolation to
the wavelengths larger than the box leads to some error in angular power
spectrum. This is further
investigated
in figure \ref{fig3_3}, where the logarithmic contribution to $C_l$'s as a
function of wavenumber $k$ (figure \ref{fig3_3}a) and redshift $z$ (figure
\ref{fig3_3}b)
is shown for several values of $l$. For low
$l$ there is a discontinuity at $k=0.27h$Mpc$^{-1}$
caused by a poor matching of the two power spectra at late times.
This discontinuity is
the largest for values of $l$ which receive the dominant contribution from the
wavelengths around the box size at late epochs ($z<1$).
The uncertainty in $C_l$ because of this is at most
20-30\% for $l \approx 100$ and is significantly smaller
at larger $l$.
%For intermediate $l$ the dominant uncertainty arises from using
%only second order results for $z>9$, where
%dominant one must use the results of
%N-body simulation to obtain accurate results even at early times,
%This causes
%another discontinuity, which can be seen in figure \ref{fig3_3} at $l=3000$.
%This effect is less important and leads to at most a
%few percent uncertainty in $C_l$'s.
For large $l$ the small scale cut-off limits the angular resolution of the
RS effect.
In principle the integral in equation \ref{phidot} should be performed from the
observer to the last-scattering surface, however due to the finite resolution
in the N-body simulation one can only start to integrate from
$r=l/k_{max}$, where in the present case
$k_{max}=25h$Mpc$^{-1}$. For low $l$ this results in a
few Mpc cutoff in $r$, rising up to 400$h^{-1}$
Mpc at $l=10000$.
At this value of $l$
the dominant scale is $k \approx 5h$Mpc$^{-1}$
(figure \ref{fig3_3}a) and the scales with $k>25h$Mpc$^{-1}$ still
have a negligible
contribution. Only for $l\gg 10000$ the scales smaller than (25h)$^{-1}$Mpc
become significant and limit the dynamic range of angular power spectrum.
The redshift distribution studied in figure \ref{fig3_3}b
indicates that
the typical contribution to the RS effect comes from $z$ around 1 at
$l>1000$.
At lower $l$ the dominant
contribution comes from redshifts below 1
(e.g. $z \approx 0.2$ at $l \approx 100$),
but is never dominated by very nearby structures. This is comforting as it
guarantees that the fair sampling criterion is satisfied. Moreover, the
observational bias caused by observing areas of the sky which do not contain
large nearby clusters
should be small for all but the smallest values of $l$.

The conclusion derived from this paper is that
in the models with
no early reionization the Rees-Sciama
effect is negligible compared to the primary anisotropies on
all observationally interesting scales ($\theta>1'$)
and is in any case below the
present-day observational limits ($\sim 10^{-6}$ in $\Delta T/T$)
on all angular scales.
While the results presented here are specific to the
flat CDM models, other models that reproduce the observed cluster abundance and
large-scale correlations should give comparable results. The main additional
effect present in the
models with $\Omega_{m0} < 1$ is the decay of potential on
linear scales (equations \ref{phidot}-\ref{dopen}),
which gives an important additional linear
contribution to the CMB anisotropies on very
large scales (Kofman \& Starobinski 1985; Kamionkowski \& Spergel 1994)
and also on smaller scales in reionized models (Hu \& Sugiyama 1994).
The nonlinear RS effect itself actually decreases in low $\Omega_{m0}$ models
because for a given density normalization both the linear potential
and the linear velocity decrease with $\Omega_{m0}$ and lead to a smaller
$\dot{\phi}$ (this will be partially offset by the
longer comoving radial pathlength).
The linear RS effect is also present
in mixed or hot dark matter models, where massive neutrinos contribute
to the dark matter and their free-streaming causes the potential to
change in time at late epochs. This leads to a small, but potentially
measurable effect on primary CMB anisotropies (e.g. Ma \&
Bertschinger 1995).
If the universe was
reionized early enough so that it became optically thick,
then the primary anisotropies would have been erased and
the RS effect would dominate over the primary contribution
at a much lower $l$. However, in this case
secondary anisotropies caused by the Vishniac effect would
also be more important
and would swamp the RS effect, as they give few times $10^{-6}$
contribution in the early reionized universe (e.g. Persi et al. 1995).
Therefore,
the Rees-Sciama effect is likely to be unimportant
on arcminute scales and larger regardless of the particular model of structure
formation or of its reionization history.

\acknowledgements
I would like to thank Ed Bertschinger for providing the N-body
data and for many useful discussions, Bhuvnesh Jain for
insights on the second order perturbation theory calculation and both
for a careful reading of the manuscript. This
work was supported by NSF grant AST-9318185 and NASA grant NAG5-2186.
Supercomputing time was generously provided by the Cornell Theory Center
and the National Center for Supercomputing Applications.

\begin{figure}[p]
\vspace*{8.0 cm}
\caption{Comparison between various power spectra discussed in the text at
$z=4$. MSS denotes the Mart\' inez-Gonzalez et al. (1994) approximation using
the evolution of the density power spectrum alone. The curve was computed by
finite differencing of two power spectra at different times and is noisier
than other spectra, which are computed at the same time.}
\includegraphics{fig3_1.ps}

\label{fig3_1}
\end{figure}

\begin{figure}[p]
\vspace*{7.5 cm}
\caption{Comparison between N-body and second order calculation of $\delta +
i\eta^{-1}\vec k \cdot \vec p$ as a function of
expansion factor $a=(1+z)^{-1}$.
{}From bottom to top the three spectra are for $a=0.2$,
$a=0.4$ and
$a=0.8$, respectively.}
\includegraphics{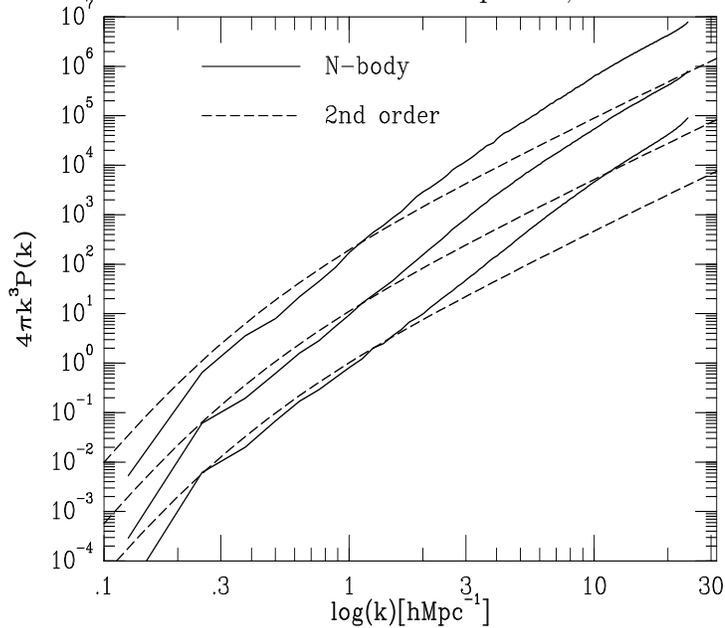}

\label{fig3_2}
\end{figure}

\begin{figure}[p]
\vspace*{5.5 cm}
\caption{RS contribution to the
angular power spectra $l(l+1)C_l/2\pi$ for various CDM models. Also
plotted is the RS effect for the standard CDM case from the second order
calculation and the primary contribution to the spectrum for a COBE normalized
adiabatic CDM model
($h=0.5$, $\Omega_{b0}h^2=0.05$), adopted from Bode \& Bertschinger (1995).}
\includegraphics{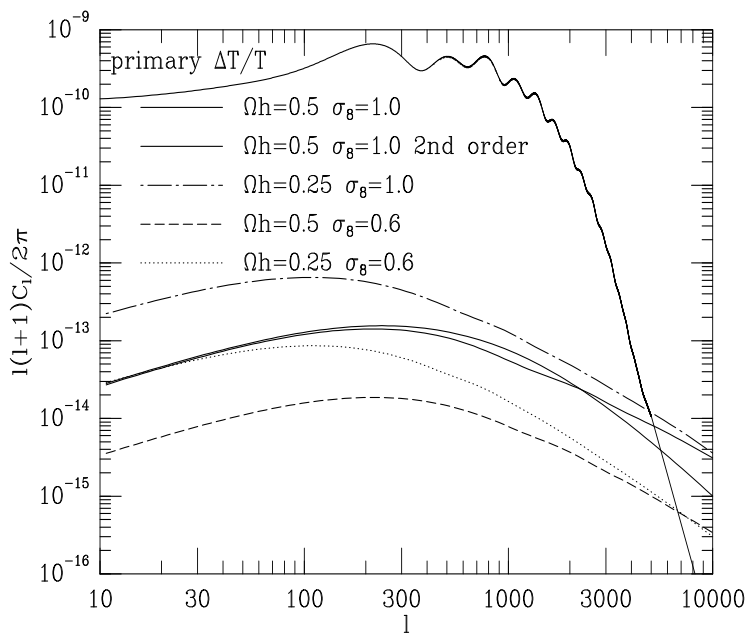}

\label{fig3_4}
\end{figure}

\begin{figure}[p]

\includegraphics{fig3_3a.ps}
\includegraphics{fig3_3b.ps}
\vspace*{17 cm}
\caption{In (a), logarithmic contribution to $C_l$'s is plotted as a
function of wavenumber $k$
for the standard CDM model.
{}From left to right the $l$ values are: 300, 1000,
3000 and 10000. In (b), logarithmic contribution to $C_l$'s as a
function of redshift $z$ is plotted for the same values of $l$.}
\label{fig3_3}
\end{figure}


\begin{thebibliography}{}
\bibitem{afs94} \reference
Arnau, J. V., Fullana, M. J., \& S\' aez, D. 1994, \mnras, 268, L17
\bibitem{bert95} \reference
Bertschinger, E. in "Cosmology and Large Scale Structure",
Les Houches summer school,
ed. R. Schaeffer, to be published (Elsevier Science
Publishers, Netherlands)
\bibitem{bertbode} \reference
Bode, P., \& Bertschinger, E. 1995, preprint astro-ph 9504040
\bibitem{bond} \reference
Bond, J. R. 1995, in "Cosmology and Large Scale Structure",
Les Houches summer school,
ed. R. Schaeffer, to be published (Elsevier Science
Publishers, Netherlands)
\bibitem{chor92} \reference
Chodorowski, M. J. 1992, \mnras, 259, 218
\bibitem{chor94} \reference
Chodorowski, M. J. 1994, \mnras, 266, 897
\bibitem{cfa} \reference
da Costa, L. N., Vogeley, M. S., Geller, M. J., Huchra, J. P., \&
Park, C. 1994, \apj 437, L1
\bibitem{gelb} \reference
Gelb, J., \& Bertschinger, E. 1994, \apj, 436, 467
\bibitem{hamilton} \reference
Hamilton, A. J. S., Kumar, P., Lu, E., \& Matthews, A. 1991, \apj, 374, L1
\bibitem{heath} \reference Heath, D. J., 1977, \mnras, 179, 351
\bibitem{husuga} \reference
Hu, W., \& Sugiyama, N. 1994, Phys. Rev. D, 50, 627
\bibitem{husugb} \reference
Hu, W., \& Sugiyama, N. 1995, Phys. Rev. D, 51, 2599
\bibitem{jain} \reference Jain, B., \& Bertschinger, E., 1994, \apj, 431, 495
\bibitem{kaiser82} \reference
Kaiser, N. 1982, \mnras, 198, 1033
\bibitem{kamio} \reference
Kamionkowski, M. \& Spergel, D. N. 1994, \apj, 432, 7
\bibitem{kofman} \reference
Kofman, L., \& Starobinsky, A. 1985, Sov. Astron. Lett., 11, 271
\bibitem{ma} \reference
Ma, C. P., \& Bertschinger, E. 1995, \apj, to be published
\bibitem{mg90} \reference
Mart\' inez-Gonz\' alez, E., Sanz, J. L., \& Silk, J. 1990, \apj, 355, L5
\bibitem{mg92} \reference
Mart\' inez-Gonz\' alez, E., Sanz, J. L., \& Silk, J. 1992, Phys. Rev. D, 46,
4196
\bibitem{mg94} \reference
Mart\' inez-Gonz\' alez, E., Sanz, J. L., \& Silk, J. 1994, \apj, 436, 1
\bibitem{Mollerach} \reference
Mollerach, S., Gangui, A., Lucchin, F., \& Matarrese, S. 1995, preprint
\bibitem{munshi} \reference
Munshi, D., Souradeep, T., \& Starobinski, A. A. 1994, preprint IUCAA-38/94
\bibitem{nottale} \reference
Nottale, L. 1984, \mnras, 206, 713
\bibitem{panek} \reference
Panek, M. 1992, \apj, 388, 225
\bibitem{peacock} \reference
Peacock, J. A., \& Dodds, S. J. 1994, \mnras, 267, 1020
\bibitem{peebles} \reference
Peebles, P. J. E., The Large Scale Structure of the Universe (Princeton
University, Princeton, NJ, 1980)
\bibitem{persi} \reference
Persi, F. M., Spergel, D. N., Cen, R., \& Ostriker, J. P. 1995, \apj,
442, 1
\bibitem{rs} \reference
Rees, M. J., \& Sciama, D. W. 1968, Nature, 517, 611
\bibitem{sachs} \reference Sachs, R. K., \& Wolfe, A. M. 1967, \apj, 147, 73
\bibitem{seljak94} \reference
Seljak, U. 1994, \apj, 435, L87
\bibitem{seljak95} \reference
Seljak, U. 1995, \apj, submitted
\bibitem{Smoot92} \reference
Smoot, G. F. et al. 1993, \apj, 396, L1
\bibitem{tv} \reference
Thompson, K. L., Vishniac, E. T. 1987, \apj, 313, 517
\bibitem{tuluie} \reference Tuluie, R., \& Laguna, P. 1995, \apj, 445, L73
\bibitem{white} \reference
White, S. D. M., Efstathiou, G., \& Frenk C. S. 1993, \mnras, 262, 1023
\end{thebibliography}
\end{document}